\documentclass[manuscript]{aastex}

\shorttitle{Radiative Transfer of synchrotron radiation}
\shortauthors{Cawthorne and Hughes.}

\begin{document}

\title{The radiative transfer of synchrotron radiation through a compressed random
magnetic field.}

\author{T. V. Cawthorne}
\affil{Jeremiah Horrocks Institute, University of Central Lancashire,
    Preston, Lancashire, PR1 2HE, U.K.}
\and
\author{P. A. Hughes}
\affil{Department of Astronomy, University of Michigan, Ann Arbor, MI 48109-1042,  U.S.A.}

\renewcommand{\baselinestretch}{1.5} 
\begin{abstract}

This paper examines the radiative transfer of synchrotron radiation in the presence of
a magnetic field configuration resulting from the compression of a highly disordered 
magnetic field. It is shown that,
provided
Faraday rotation and circular polarization can be neglected, the radiative transfer
equations for synchrotron radiation separate for this configuration, and the 
intensities and polarization values
for sources that are uniform on large scales can be found straightforwardly
in the case where opacity is significant. 
Although the emission and absorption coefficients must, in general, be obtained 
numerically, the process is much simpler than a full numerical solution to the 
transfer equations. Some illustrative results are given and an interesting effect,
whereby the polarization increases while the magnetic field distribution becomes
less strongly confined to the plane of compression, is discussed.

The results are of importance for the interpretation of polarization near the edges of
lobes in radio galaxies and of bright features in the parsec--scale jets of AGN, where such 
magnetic field configurations are believed to exist. 

(Note: The original ApJ version of this paper contained two errata, which are 
corrected in this version of the paper. First, Fig. 2 was plotted 
as a mirror image of the correct version, reflected
about the line $\log_{10}(\nu/ \nu_0)=0$. Second, as stated, Equation A16 is valid for
$\gamma=2$, not $\gamma=3$.)

\end{abstract}


\section{Introduction.}
Models for relativistic jet production in active galactic nuclei
strongly favor ordered magnetic fields within thousands of gravitational
radii of the central supermassive black hole 
\citep{2012SSRv..169...27P}  even when complex flows and instability
are admitted \citep{2012MNRAS.423.3083M,2013MNRAS.429.2482P}. Such fields
are often assumed to persist to the parsec and even kiloparsec scale, and
indeed might be required to explain observations as diverse as transverse
gradients in Faraday rotation measure \citep{2012SSRv..169...27P} and the
extraordinary stability of flows such as that revealed by radio and X-ray
observations of Pictor~A \citep{2001ApJ...547..740W}.

Nevertheless, compelling evidence exists that a substantial fraction of
the magnetic field energy is in a random component, from the sub-parsec
to kiloparsec scales. Following an analysis of cm-band single-dish
data by \citet{1985ApJ...290..627J}, which revealed a magnetic field
structure capable of explaining the ``rotator events'' seen in time
series data of Stokes parameters $Q$ and $U$, activity in a number of
AGN has been successfully modeled by shocks that compress an initially
tangled magnetic field, increasing the percentage polarization during
outburst \citep{1989ApJ...341...68H, 1991ApJ...374...57H}. Such
a model has recently been extended to incorporate oblique shocks
\citep{2011ApJ...735...81H}.  Tangled magnetic fields carried through
conical shock structures have been explored by \citet{2006MNRAS.367..851C},
and this picture has been successfully used to explain the characteristics
of a stationary jet feature in 3C~120 \citep{2012ApJ...752...92A} and
has been suggested as an explanation of multiwavelength variations of
3C~454.3 \citep{2012ApJ...758...72W}. On the larger (kiloparsec) scale,
\citet{2002MNRAS.336..328L} have pioneered analysis of the magnetic field
structure of jets, most recently concluding \citep{2006MNRAS.372..510L}
that the jet in 3C~296 has a random but anisotropic magnetic field structure.

The spectral, spatial, and temporal behavior of the Stokes parameters $Q$
and $U$ provides a powerful diagnostic of the magnetic field structure,
and thus indirectly, of the flow character in such jets, and the degree
of linear polarization for a compressed, tangled magnetic field (due,
for example, to a shock) was explored in the optically thin limit by
\citet{1985ApJ...298..301H}. 
(An earlier paper, \citet{1980MNRAS.193..439L}, also considered 
this kind of structure in the limit of an infinitely strong compression.)
However, at least on the parsec and sub-parsec
scale these flows exhibit opacity. Indeed, the ``core'' seen in 
low frequency ($\nu \le 10$\,GHz) VLBI maps
is widely interpreted as being the ``$\tau=1$-surface'': the location of
the transition from optically thin to optically thick emission at the observing
frequency of the map \citep{2006AIPC..856....1M}. This location within the
jet has a special significance for jet studies, as it is by definition the
surface from which propagating components first appear as distinct features
on the map; there is compelling evidence that $\gamma$-ray flares arise
close to the mm-wave core \citep{2010ApJ...710L.126M}, understanding the
origin of which requires knowledge of the flow conditions there. At 
these higher frequencies, the
interpretation of the core as the $\tau=1$-surface is certainly complicated
by the presence of stationary features (possibly recollimation shocks)
which, even if responsible for the core in some sources, must lie close
to regions of significant opacity \citep{2006AIPC..856....1M}. It would
therefore be of great value to have a description of the polarized emission
from compressed, tangled magnetic fields in the presence of opacity.

In earlier work,
\citet{1986A&A...164L..16C,1988A&A...196..327C} calculated the emitted
synchrotron intensity (Stokes I) for a purely random magnetic field.
\citet{1990ApJ...360..417C} extended the discussion to polarization (Stokes
I, Q \& U) but the considered field geometry remained purely random,
with zero mean polarization, the focus of the study being observable
fluctuations -- in particular the rms deviations from the means -- for
various models of the magnetic field turbulent structure, with finite
coherence scale and a finite number of magnetic cells.  In a recent major
study \citet{2012ApJ...747....5L} admitted a mean magnetic field, but
were concerned with axisymmetric turbulence that leads to anisotropic
intensity (Stokes I) fluctuations.  The primary goal was to facilitate
probing Galactic MHD turbulence.  Our work, on the other hand, considers
a compressed random field, such as would result from a shock, or subsonic
disturbance, leading to non-zero mean polarization, and computes that
in the limit $N_{\rm cells}\rightarrow\infty$, so that only the smooth,
mean behavior is described.

For a parsec-scale jet magnetic field of $10^{-5}$T
\citep{2009MNRAS.400...26O} the gyroradius of an electron with $\gamma=10^2$
is $1.7\times 10^4$m, more than twelve orders of magnitude less than
the system scale, thus permitting a very small scale turbulent field --
effectively an infinite number of cells within a telescope beam -- without
the cell scale approaching the gyroradius.  Typical radio source hotspot
fields are $10^{-9}$T \citep{2003ApJ...584..643D}; thus the ratio of
system scale (kpc) to gyroradius is only one order of magnitude less for
particles of the same energy, and no more than three orders of magnitude
less for particles of an energy radiating in the same radio waveband. This
still comfortably permits a small scale turbulent field with effectively
an infinite number of cells within the volume without the cell scale
approaching the gyroradius. In no case would our approximation make it
necessary to consider turbulent cell sizes so small that jitter radiation
\citep{2000ApJ...540..704M,2006MNRAS.365L..11F} is important.

\section{Propagation of synchrotron radiation through a compressed random field.}

This section demonstrates that the radiative transfer equations for the propagation of synchrotron
radiation through a
compressed random field separate, provided circular polarization and Faraday rotation can be neglected. 
The resulting absorption and emission coefficients are obtained in Appendix A. 
The approach follows those of
Appendix A in Hughes, Aller \& Aller (1985) and Chapter 3 from Pacholczyk (1970).
In order to obtain consistency between these two works, the coordinate system used in Hughes, Aller \& Aller
have been relabelled as follows.

\begin{figure}
\begin{center}
\includegraphics[angle=0,height=6.5cm]
{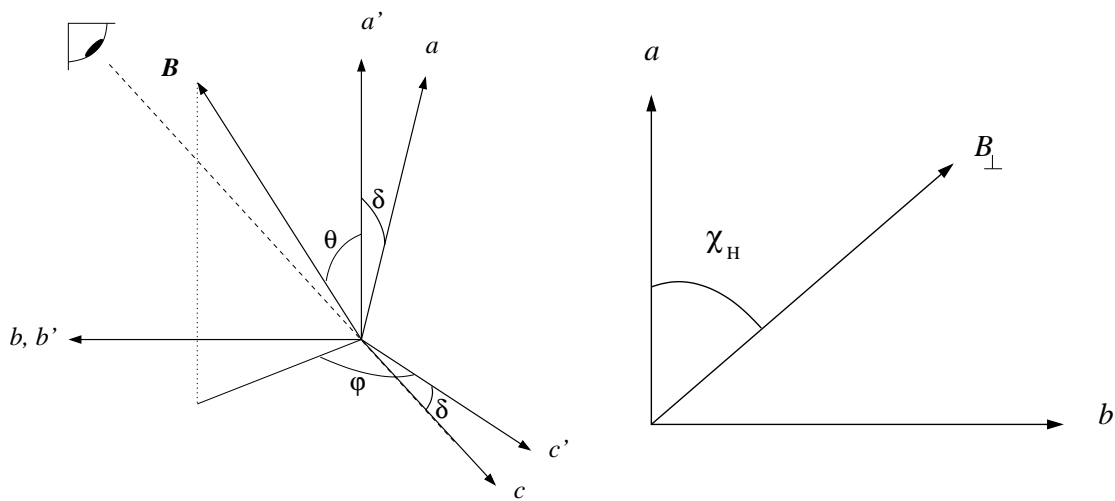}
\caption {\label{comb} Left diagram: This figure illustrates the coordinate systems used in this paper. Both the $a$ and $a'$ axes
and the $c$ and $c'$ axes are inclined at angle $\delta$.  The $b$ and $b'$ axes are coincident. 
The magnetic field is defined with respect to the $(a',b',c')$ coordinate system.
Plasma with disordered magnetic field is compressed along the direction
parallel to the $a'$ axis. Radiation is observed propagating along the $-c$ axis. Right diagram:
This figure illustrates the sky plane, with the $c$ axis pointing away from the observer. $\chi_H$ is the angle
between the $a$ axis and the projection of the magnetic field onto the sky plane.}
\end{center}
\end{figure}

Before compression, the direction of the magnetic field is defined by reference to the $(a',b',c')$ 
coordinate system as shown in Fig.~\ref{comb} (left panel). The polar angle $\theta$ separates the $a'$ axis and the direction
of the local magnetic field, and
the azimuthal angle $\phi$ separates the $c'$ axis and the projection of the field 
onto the $b',c'$ plane. In this system,
$B_{a'}  =  B_0 \cos\theta$, 
$B_{b'} =  B_0 \sin\theta\sin\phi $ 
and $B_{c'} = B_0 \sin\theta\cos\phi$. 
After compression, such that unit length parallel to the $a'$ axis is reduced to length $K$, 
the requirement that magnetic flux is conserved yields a magnetic field with components 
$B_{a'}  =  B_0 \cos\theta$, 
$B_{b'} =  B_0 \sin\theta\sin\phi/K$ and 
$B_{c'} = B_0 \sin\theta\cos\phi /K$. 
A rotation of the coordinate system through angle $\delta$ about the $b'$ axis gives the $a,b,c$ coordinate system
(chosen so that the observer lies
on the $-c$ axes) in terms of which the local magnetic field is
\begin{eqnarray}
B_{a} & = & B_0 (\cos\theta\cos\delta + \sin\theta\cos\phi\sin\delta/K) \label{ba}  \\
B_{b} &= & B_0 \sin\theta\sin\phi/K  \label{bb} \\ 
B_{c} &=& B_0 (\sin\theta\cos\phi \cos\delta/K - \cos\theta\sin\delta) \label{bc}
\end{eqnarray}

These results can be obtained from Hughes, Aller \& Aller (1985) by making the substitutions
($x\rightarrow -c'$, $y\rightarrow b'$, $z \rightarrow a'$, $x'\rightarrow -c$, $y' \rightarrow b$,
$z' \rightarrow a$, and $\epsilon \rightarrow -\delta$). 

Following Pacholczyk (1970) Equation 3.66 and assuming that the circular polarization and Faraday rotation are
negligible, the radiative transfer equations are written in terms of 
$I^{(a)}$ and $I^{(b)}$, the intensities measured by dipoles aligned with the $a$ and $b$ axes, 
respectively, and the Stokes parameter $U^{(ab)}$:
\begin{eqnarray}
\frac{dI^{(a)}}{ds} &=& I^{(a)}[-\kappa^{(1)} \sin^4\chi_H -\kappa^{(2)} \cos^4\chi_H - \frac{1}{2}\kappa\sin^2 2\chi_H]  \nonumber \\
                  &+& U^{(ab)}[\frac{1}{4}(\kappa^{(1)}-\kappa^{(2)})\sin 2\chi_H]
                  + \epsilon^{(1)}\sin^2\chi_H + \epsilon^{(2)}\cos^2\chi_H  \label{rt1} \\
\frac{dI^{(b)}}{ds} &=& I^{(b)}[-\kappa^{(1)} \cos^4\chi_H -\kappa^{(2)} \sin^4\chi_H - \frac{1}{2}\kappa\sin^2 2\chi_H]  \nonumber \\
                  &+& U^{(ab)}[\frac{1}{4}(\kappa^{(1)}-\kappa^{(2)})\sin 2\chi_H] 
                  + \epsilon^{(1)}\cos^2\chi_H + \epsilon^{(2)}\sin^2\chi_H \label{rt2} \\
\frac{dU^{(ab)}}{ds} &=& (I^{(a)}+I^{(b)})\frac{1}{2}(\kappa^{(1)}-\kappa^{(2)})\sin\,2\chi_H  
                    -\kappa U^{(ab)} - (\epsilon^{(1)}-\epsilon^{(2)})\sin\,2\chi_H \label{rt3}
\end{eqnarray}
Here,  $\chi_H$ is the angle between the $a$ axis and the projection of the magnetic field onto the plane of the sky, as shown in 
Fig.~\ref{comb} (right diagram).
$\kappa^{(1)}$ and $\kappa^{(2)}$ are, respectively, the absorption coefficients for planes of (electric field) polarization 
perpendicular and parallel to the projected magnetic field. Likewise,
$\epsilon^{(1)}$ and $\epsilon^{(2)}$ are, respectively, the emission coefficients for planes of (electric field) polarization 
perpendicular and parallel to the projected magnetic field. The polarization--averaged absorption coefficient is defined by
\begin{eqnarray}
\kappa=(\kappa^{(1)}+\kappa^{(2)})/2 \label{avabsco}
\end{eqnarray}

For a power--law distribution of radiating electrons such that the density of electrons in the energy interval
$dE$ is $\mathcal{N}(E)dE=\mathcal{N}_0 E^{-\gamma}dE$,
the emission and absorption coefficients for a region with uniform field are given by Pacholczyk (1970) as 
\begin{eqnarray}
\epsilon^{(1),(2)} &=&  C\mathcal{N}_0 B_{\perp}^{(1+\gamma)/2} \nu^{(1-\gamma)/2} \left[1 \pm \frac{\gamma+1}{\gamma+7/3}\right]  \label{emco} \\
\kappa^{(1),(2)} &=&   D\mathcal{N}_0 B_{\perp}^{(2+\gamma)/2} \nu^{-(4+\gamma)/2} \left[1 \pm \frac{\gamma+2}{\gamma + 10/3} \right] \label{absco}
\end{eqnarray}
where the constants $C$ and $D$ are given in Appendix B. Inside the square brackets, the plus sign refers to polarization (1)
and the minus sign to polarization (2).  

\subsection{Separation of the transfer equations.}

 Equations~\ref{rt1} to \ref{rt3} contain a term describing the contribution to
$I^{(a),(b)}$ and $U^{(ab)}$ due to polarized absorption, which depends on
$(\kappa^{(1)}-\kappa^{(2)})\sin 2\chi_H$. For the power-law distribution of
particles considered here it is always true that $\kappa^{(1)}>\kappa^{(2)}$,
Thus in the $(12)$ frame, these
contributions are always in the same sense. For a uniform magnetic field
this term is zero because polarized
absorption orthogonal to the field does not contribute to the mode parallel
to the field, and vice versa. If
we consider a partially compressed random magnetic field as equivalent
to the sum of a uniform component orthogonal to the sense of compression,
plus a superposed random distribution of field elements, the latter will
not reintroduce contributions from these difference terms, as their random
distribution guarantees that they do not modify the polarized component of
the radiation. Equivalently, while a compressed magnetic field exhibits a
preferred sense -- the plane of compression -- and the projected magnetic
field elements will be distributed with a narrow dispersion in $\chi_H$
about the projection of this direction on the plane of the sky, on average
there will be as many elements with $\chi_H>0$ as there are with $\chi_H<0$,
with no net effect upon the radiation field. 
A more formal demonstration of this result follows. 

From Equations~\ref{rt1} and \ref{rt2},
\begin{eqnarray}
\kappa^{(1)}-\kappa^{(2)} &=& 
   D\mathcal{N}_0 B_{\perp}^{(2+\gamma)/2} \nu^{-(4+\gamma)/2} \left[\frac{2(\gamma+2)}{\gamma + 10/3} \right] \nonumber \\
         &=& D\mathcal{N}_0 (B_{a}^{2}+B_{b}^{2})^{(2+\gamma)/4} \nu^{-(4+\gamma)/2} \left[\frac{2(\gamma+2)}{\gamma + 10/3} \right] \label{kdif}
\end{eqnarray}
From Fig.~\ref{comb} (right diagram),
\begin{eqnarray}
\sin \chi_H &=& \frac{B_b}{(B_{a}^{2} + B_b^{2})^{1/2}} \label{sinchi} \\
\cos \chi_H &=& \frac{B_a}{(B_{a}^{2} + B_b^{2})^{1/2}} \label{coschi}
\end{eqnarray}
and so
\begin{eqnarray}
\sin\,2\chi_H 
              &=& 2 B_a B_b/(B_{a}^{2} + B_{b}^{2}) \label{sin2chi} 
\end{eqnarray}
so that
$(\kappa^{(1)}-\kappa^{(2)})\sin\,2\chi_H \propto B_a B_b(B_{a}^{2}+B_{b}^{2})^{(\gamma-2)/4}$. 
Averaging this expression over all $\theta$ and $\phi$ gives 
\begin{eqnarray}
<(\kappa^{(1)}-\kappa^{(2)})\sin\,2\chi_H > \propto \frac{1}{4\pi} \int_{-\pi}^{\pi} \int_{0}^{\pi} B_a B_b(B_{a}^{2}+B_{b}^{2})^{(\gamma-2)/4} 
                               \sin\theta\, d\theta \, d\phi \label{avkdif}
\end{eqnarray}
From Equations~\ref{ba} to \ref{bc}, it is clear that $B_a$ is an even function of $\phi$ while $B_b$ is an odd function of $\phi$, so
that the integrand in Equation~\ref{avkdif} is an odd function of $\phi$. Integrating with respect to $\phi$ from $-\pi$ to $\pi$ therefore
yields the result zero. Therefore
\begin{eqnarray}
<(\kappa^{(1)}-\kappa^{(2)})\sin\,2\chi_H> &=& 0 \label{kdif0}
\end{eqnarray}
Hence, $dI^{(a)}/ds$ depends only
on $I^{(a)}$, $dI^{(b)}/ds$ depends only on $I^{(b)}$ and $dU^{(ab)}/ds$ depends only on $U^{(ab)}$. In this case, the equations
separate and have straightforward solutions. 

Very similar similar arguments apply to the term $(\epsilon^{(1)}-\epsilon^{(2)})\sin 2\chi$
in Equation~\ref{rt3}, which describes the contribution
of polarized emission to $U^{(ab)}$, so that $<(\epsilon^{(1)}-\epsilon^{(2)})\sin\,2\chi_H> = 0$.

Assuming that the magnetic field is disordered on scales small compared those over which the
radiation field changes significantly,
the emission and absorption coefficients can be averaged over initial magnetic field direction and 
the radiative transfer Equations~\ref{rt1} to \ref{rt3} thus simplify to
\begin{eqnarray}
\frac{dI^{(a)}}{ds} &=& -<\kappa^{(a)}> I^{(a)} + <\epsilon^{(a)}> \label{rts1} \\
\frac{dI^{(b)}}{ds} &=& -<\kappa^{(b)}> I^{(b)} + <\epsilon^{(b)}> \label{rts2} \\
\frac{dU^{(ab)}}{ds} &=& -<\kappa> U^{(ab)} \label{rts3}
\end{eqnarray}
for which, in a uniform source, the following solutions can be obtained straightforwardly:
\begin{eqnarray}
I^{(a)}(s) &=& \frac{<\epsilon^{(a)}>}{<\kappa^{(a)}>}(1-\exp(-<\kappa^{(a)}>s)) + I^{(a)}(s=0)\exp(-<\kappa^{(a)}>s) \label{iasol} \\
I^{(b)}(s) &=& \frac{<\epsilon^{(b)}>}{<\kappa^{(b)}>}(1-\exp(-<\kappa^{(b)}>s)) + I^{(b)}(s=0)\exp(-<\kappa^{(b)}>s) \label{ibsol} \\
U^{(ab)}(s)&=& U^{(ab)}(s=0)\exp(-<\kappa> s) \label{usol}
\end{eqnarray}
where $I^{(a)}(s=0)$, $I_b(s=0)$ and $U^{(ab)}(s=0)$, are the values incident upon the source, $s$ is the
path length through the source, and 
\begin{eqnarray}
\epsilon^{(a)} &=&  \epsilon^{(1)}\sin^2{\chi_H} + \epsilon^{(2)}\cos^2{\chi_H} \label{epsadef} \\
\epsilon^{(b)} &=&   \epsilon^{(1)}\cos^2{\chi_H} + \epsilon^{(2)}\sin^2{\chi_H} \label{epsbdef} \\
\kappa^{(a)} &=& \kappa^{(1)}\sin^{4}\chi_H + \kappa^{(2)}\cos^4\chi_H + \frac{1}{2}\kappa \sin^2 2\chi_H  \label{kadef} \\
\kappa^{(b)} &=& \kappa^{(1)}\cos^{4}\chi_H + \kappa^{(2)}\sin^4\chi_H + \frac{1}{2}\kappa \sin^2 2\chi_H  \label{kbdef}
\end{eqnarray}

\section{Results.}

Appendix A shows how the emission coefficients $<\epsilon^{(a),(b)}>$ and absorption coefficients $<\kappa^{(a),(b)}>$ can be expressed 
in terms of the function $F_{\gamma}^{(a),(b)}$ (Equations~\ref{epsaav} to \ref{fdef} and Equation~\ref{kapabav}). The integrals in
these expressions are not, in general, analytically tractable, but since $F_{\gamma=3}^{(a),(b)}$ has a simple solution 
(Equations~\ref{kapa2}, \ref{kapb2} and \ref{kapav2}), simple formulae exist for the emission coefficients when $\gamma=3$ and 
for the absorption 
coefficients when $\gamma=2$. A rough analytical approximation to $F_{\gamma=2}^{(a),(b)}$ is given by Equations~\ref{faapp} and $\ref{fbapp}$
and correction factors are plotted in Fig.~\ref{corf}. These allow computation of intensities and polarization in the case $\gamma=2$
without resort to a computer. Expressions for the constants $C$, $D$ and $\mu$ are given in Appendix B, and their values are given 
in Table\,1 for some values of $\gamma$ in the range of greatest interest. 

Results illustrating how the emergent polarization varies with frequency $\nu$ and line of sight angle $\delta$
are presented below. The integrals were performed numerically using Simpson's rule with $50$ evaluations per integral. 
Comparison between results that can be obtained analytically and the corresponding values obtained numerically suggests that
the latter are accurate four significant figures at least.

\subsection{Polarization as a function of frequency.}

If the source is uniform on scales over which the intensity changes significantly, then the solutions
given by Equations~\ref{iasol}, \ref{ibsol} and
\ref{usol} apply. It is convenient to define a characteristic frequency, $\nu_0$, at which the polarization averaged opacity is
unity, i.e.,   
\begin{eqnarray}
<\kappa>L = (<\kappa^{(1)}>+<\kappa^{(2)}>)L/2 = 1  \label{op1} 
\end{eqnarray}
(Note that $\nu_0$ will be a function of $K$, $\delta$ and $\gamma$.)
Then, from Equations~\ref{kapabav} and \ref{kaptav}, the opacities in polarizations $a$ and $b$ are 
\begin{eqnarray}
\tau^{(a),(b)} &=& <\kappa^{(a),(b)}>L = \left(\frac{\nu}{\nu_0}\right)^{-(\gamma+4)/2}\frac{F_{(\gamma+1)}^{(a),(b)}(\delta,K)}{H_\gamma(\delta,K)} 
\label{tauab}
\end{eqnarray}
\begin{figure}
\begin{center}
\includegraphics[angle=-90,width=13cm]
{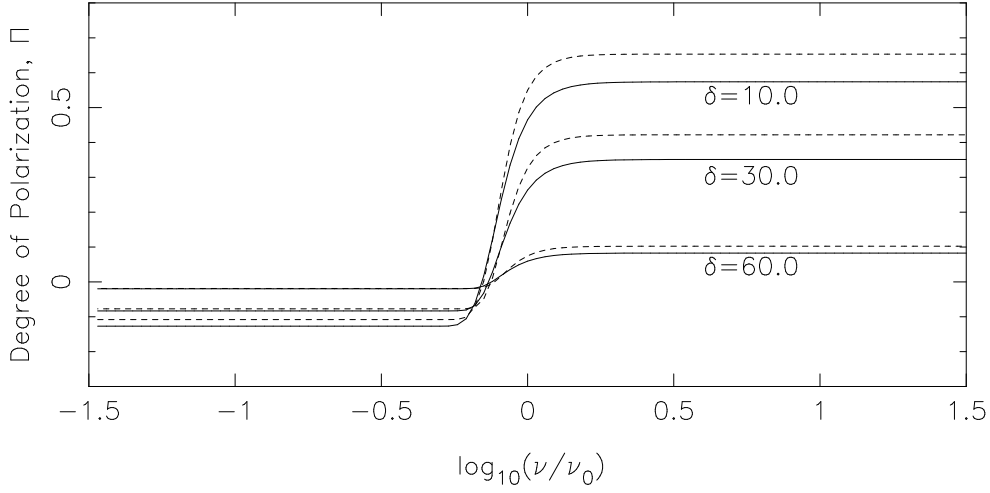}
\caption {\label{pivnu} The degree of polarization as a function of frequency for three values of the inclination angle, $\delta$. The 
continuous lines are for $\gamma=2$, the dashed lines for $\gamma=3$. The compression factor is $K=0.2$.}
\end{center}
\end{figure}
The intensities are then given by Equations~\ref{iasol}, ~\ref{ibsol} and \ref{epskap} 
\begin{eqnarray}
I^{(a),(b)} &=& \frac{\mu m \nu^{5/2}}{\nu_L^{1/2}}\frac{F_{\gamma}^{(a),(b)}}{F_{\gamma+1}^{(a),(b)}}\left(1 - e^{-\tau^{(a),(b)}}\right) \label{iab}
\end{eqnarray}
and the degree of polarization is then
\begin{eqnarray}
\Pi = \frac{I^{(a)}-I^{(b)}}{I^{(a)}+I^{(b)}} \label{dopdef}
\end{eqnarray}
The spectral variation of the degree of polarization is illustrated in Fig.~\ref{pivnu} for three values of $\delta$, two
values of $\gamma$, and $K=0.2$. The figure illustrates the transition from optically thin emission, where the polarization
fraction is generally high and the ($E$ field) polarization direction is parallel to the $a$ axis $(\Pi >0)$, to optically thick 
emission, where
the polarization is generally lower, and the polarization direction is parallel to the $b$ axis $(\Pi < 0)$.
The polarization decreases as $\delta$, the angle of inclination between the line of sight and the plane of compression,
increases, and the disordered component of the magnetic field becomes more apparent. 

In the optically thin (high frequency) limit, the degree of polarization is
\begin{eqnarray}
\Pi_{thin} &=& \frac{F_\gamma^{a}-F_\gamma^{(b)}}{F_\gamma^{(a)}+F_\gamma^{(b)}}  \label{pithin}
\end{eqnarray}
while in the optically thick (low frequency) limit, the degree of polarization is
\begin{eqnarray}
\Pi_{thick} &=& \frac{(F_\gamma^{(a)}/F_{\gamma+1}^{(a)})-(F_\gamma^{(b)}/F_{\gamma+1}^{(b)})}{(F_\gamma^{(a)}/F_{\gamma+1}^{(a)})+
                (F_\gamma^{(b)}/F_{\gamma+1}^{(b)})}  \label{pithick}
\end{eqnarray}

\begin{figure}
\begin{center}
\includegraphics[angle=-90,width=12cm]
{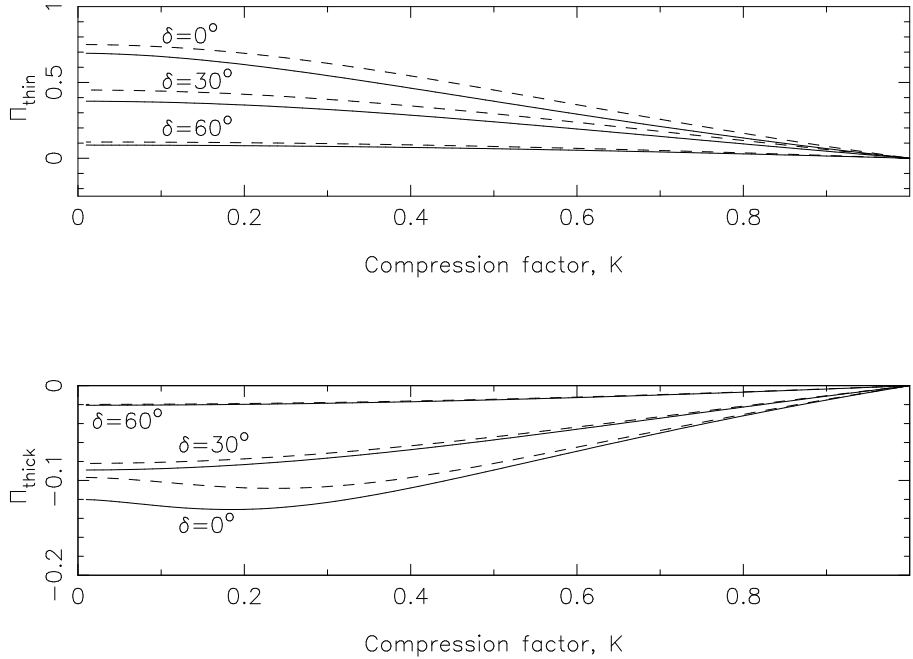}
\caption {\label{pivk} The degrees of polarization in the optically thin limit (above) and the optically thick
limit (below) are plotted as a function of $K$ for the values of $\delta$ shown. The continuous lines show results for
$\gamma=2$, the dashed lines for $\gamma=3$.}
\end{center}
\end{figure}

These values are plotted as a function of compression factor $K$, for various values of the inclination angle $\delta$, in 
Fig~\ref{pivk}. 
As expected, in the optically thin limit, the degree of polarization decreases monotonically with increasing $K$, and with
increasing $\delta$.  
The value of $\Pi$ in the optically thick limit generally
decreases in magnitude as $K$ increases, though for $\delta$ less than about $10^{\circ}$, the value of $\Pi_{thick}$ has a turning point at about
$K=0.2$. It is, at first sight, surprising that, as $K$ increases from zero and the field becomes more isotropic (or less strongly confined 
to the plane of compression), the degree of polarization actually increases. This occurs because, although both emission and absorption
coefficients for the two polarizations become closer, as clearly they should, the values of $\epsilon/ \kappa$ for the two polarizations 
initially diverge. The reason is that while $K$ is very small and increasing, both $\epsilon^{(a)}/ \epsilon^{(b)}$ and 
$\kappa^{(a)}/ \kappa^{(b)}$ decrease, but the ratio of the $\epsilon$ values decreases more strongly than that of the $\kappa$ values.
This occurs because the contribution to the coefficients from the component of field perpendicular to the plane of compression
(which, in relative terms, is increasing) is greater for the emission coefficients than the absorption coefficients, because
the latter depend more sensitively on magnetic field. This subtle effect can be more easily understood with reference to a similar but simpler
magnetic field geometry, as shown in Appendix C.  

\subsection{Polarization as a function of inclination.}

\begin{figure}
\begin{center}
\includegraphics[angle=-90,width=13cm]{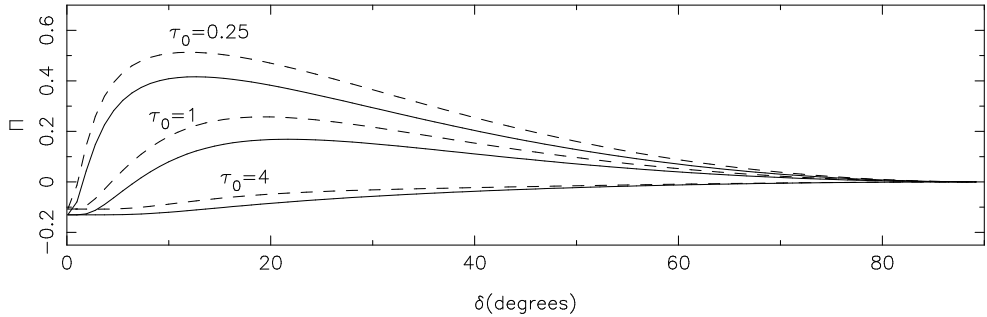}
\caption{\label{pivdelta} The degree of polarization is plotted as a function of $\delta$ for $K=0.2$ and the values of $\tau_0$ shown.
Continuous lines show results for $\gamma=2$, dashed lines for $\gamma=3$.}
\end{center}
\end{figure}

The dependence of the degree of polarization upon $\delta$, the angle of inclination between the line of sight and plane
of compression, is described below. If the emitting plasma is confined between two planes, each parallel to the plane
of compression and separated by a distance $w$, then the path length through the plasma is $L=w/ \sin\delta$. The opacity
is characterised by the value $\tau_0=<\kappa>(\delta=90^{\circ},K,\gamma)w$, the polarization averaged opacity when the line
of sight is perpendicular to the plane of compression.  The value of $\tau_0$ is given by
\begin{eqnarray}
\tau_0 &=& D\mathcal{N}_0(K)B_0^{(2+\gamma)/2}\nu^{-(4+\gamma)/2}H_{\gamma}(\delta=90^{\circ},K)w \label{tau0}
\end{eqnarray}
Then, if $\delta$ is varied while $K$, $\gamma$ and $\nu$ remain fixed,
\begin{eqnarray}
\tau^{(a),(b)} &=& \tau_0\frac{F_{\gamma+1}^{(a),(b)}(\delta,K)}{H_{\gamma}(\delta=90^{\circ},K)\sin{\delta}} \label{tauvd}
\end{eqnarray}
The intensities $I^{(a)}$ and $I^{(b)}$ and the degree of polarization are then given by Equations~\ref{iab} and \ref{dopdef}.
The results are shown in Fig.~\ref{pivdelta}, in which $\Pi$, the degree of polarization, is plotted against $\delta$, for
a compression factor $K=0.2$, and values of $\tau_0=0.25,\,1.0,\,4.0$.  
The results show that, as $\delta$ decreases from $90^{\circ}$, the polarization first rises as the partial order of the magnetic field becomes
more apparent, but then starts to fall, as opacity begins to take effect. As $\delta$ decreases
further, $\Pi$ changes from positive to negative in value (i.e. the polarization angle changes by $90^{\circ}$), and the degree
of polarization approaches the optically thick limit shown in Fig~\ref{pivk}. 
As $\tau_0$ increases in value,  the maximum (positive) value of $\Pi$ 
decreases and the frequency at which $\Pi$ changes from positive to negative increases.

\section{Summary of results.}

The radiative transfer equations for synchrotron radiation have been shown to
separate for the case of propagation through a compressed, random magnetic field,
provided Faraday rotation and circular polarization can be neglected.
Although, in general, the emission and absorption coefficients must be computed
numerically, this is still much simpler than a full numerical solution of the coupled
equations. Expressions for the emission and absorption coefficients are given in
Appendix A. Exact analytical expressions result only for the emission coefficients when
(energy index) $\gamma=3$, and for the absorption coefficients when $\gamma=2$. 
A rough approximation, together with a plot of correction factors, is given to allow
calculation of the emission coefficient for $\gamma=2$. This allows the solution
to be found for a source that is uniform on large-scales, for $\gamma=2$, 
without resort to a computer.

Some illustrative results are presented, showing the variation of polarization with 
frequency, and with inclination of the plane of compression to the line of sight. 
The optically thin and thick limits to fractional polarization are plotted against
compression factor, $K$, for 
various inclination angles. When the inclination angle $\delta<10^{\circ}$, 
the optically thick limit reveals an unusual trend in 
which, for very small $K$, the polarization increases as $K$ increases, i.e., as 
the magnetic
field becomes {\em less} strongly confined to the plane of compression. This
effect is discussed in the context of a simpler magnetic field model in Appendix C. 

\section{Acknowledgments}

TVC thanks the director of the Jeremiah Horrocks Institute at the University of 
Central Lancashire for a sabbatical semester, during which most of the
present work was undertaken. He also thanks Dr. J.-L. Gomez and the
Instituto di Astrofisica de Aldalucia in Granada for their generous 
hospitality during a part of the sabbatical. This work arose from 
discussions with Dr. Gomez during that visit. PAH was partially supported by
NASA Fermi GI grant NNX11AO13G during this work. The authors thank the
anonymous referee for a number of very useful comments on the manuscript.

(The authors also thank Mr. Christopher Kaye for identifying the two errors
in the original, published version of the paper (as described in the abstract).)
\appendix
\section{Computation of the emission and absorption coefficients.}

Following the approach of Hughes, Aller \& Aller (1985), it is convenient to define the functions $M$ and $N$ such that
\begin{eqnarray}
M(\theta,\phi) &=& (B_a/B_0)^2 
               = (\cos\theta \cos\delta + \sin\theta\cos\phi\sin\delta/K)^2 \label{mdef} \\
N(\theta,\phi) &=& (B_b/B_0)^2
                = (\sin\theta \sin\phi/K)^2 \label{ndef}
\end{eqnarray}
Furthermore, for a 1-D adiabatic compression, the particle density per unit energy, $\mathcal{N}_0=\mathcal{N}_0(K) \propto K^{-(\gamma+2)/3}$ 
(e.g., \citet{1989ApJ...341...54H}).
Then, from Equation~\ref{epsadef}, the emission coefficient $\epsilon^{(a)}$ becomes
\begin{eqnarray}
\epsilon^{(a)} &=& \epsilon^{(1)}\sin^2\chi_H + \epsilon^{(2)}\cos^2\chi_H   \nonumber  \\
               &=& C\mathcal{N}_0(K)(B_{a}{^2} +B_{b}{^2})^{(1+\gamma)/4}\nu^{(1-\gamma)/2} \left(\frac{2\gamma+10/3}{\gamma+7/3}\frac{B_b^2}{B_a^2 +B_b^2}
                                        +\frac{4/3}{\gamma+7/3}\frac{B_a^2}{B_a^2+B_b^2} \right) \nonumber \\ 
               &=& C\mathcal{N}_0(K)B_0^{(\gamma+1)/2}\nu^{(1-\gamma)/2}(M +N)^{(\gamma-3)/4}\left(\frac{(2\gamma+10/3)N + (4/3)M}{\gamma+7/3} 
                                   \right) \nonumber  \\
               &=& C\mathcal{N}_0(K)B_0^{(\gamma+1)/2}\nu^{(1-\gamma)/2}\left(\frac{\frac{4}{3}(M +N)^{(\gamma+1)/4}+2(\gamma+1)N(M+N)^{(\gamma-3)/4}}
                                                   {\gamma+7/3}\right)  \label{epsacal}
\end{eqnarray}
Similarly, from Equation~\ref{epsbdef}
\begin{eqnarray}
\epsilon^{(b)} &=& \epsilon^{(1)}\cos^2\chi_H + \epsilon^{(2)}\sin^2\chi_H   \nonumber  \\
               &=& C\mathcal{N}_0(K)B_0^{(\gamma+1)/2}\nu^{(1-\gamma)/2}\left(\frac{\frac{4}{3}(M +N)^{(\gamma+1)/4}+2(\gamma+1)M(M+N)^{(\gamma-3)/4}}
                                                   {\gamma+7/3}\right)  \label{epsbcal}
\end{eqnarray}

Averaging over the initial magnetic field direction, the emission coefficients become
\begin{eqnarray}
<\epsilon^{(a),(b)}> &=& C\mathcal{N}_0(K)B_0^{(\gamma+1)/2}\nu^{(1-\gamma)/2}F_\gamma^{(a),(b)}(\delta,K) \label{epsaav} 
\end{eqnarray}   
where
\begin{eqnarray}
F_\gamma^{(a)}(\delta,K) =  \frac{1}{4\pi}\int_{-\pi}^{\pi} \int_{0}^{\pi} 
        \frac{\frac{4}{3}(M +N)^{(\gamma+1)/4}+2(\gamma+1)N(M+N)^{(\gamma-3)/4}}{(\gamma+7/3)}\sin\theta d\theta d\phi \label{edef}
\end{eqnarray}
and
\begin{eqnarray}
F_\gamma^{(b)}(\delta,K) = \frac{1}{4\pi}\int_{-\pi}^{\pi} \int_{0}^{\pi} 
        \frac{\frac{4}{3}(M +N)^{(\gamma+1)/4}+2(\gamma+1)M(M+N)^{(\gamma-3)/4}}{(\gamma+7/3)}\sin\theta d\theta d\phi  \label{fdef}
\end{eqnarray}

The absorption coefficients are given by Equations~\ref{absco},\,\ref{kadef} and \ref{kbdef}. It is convenient to express
$\kappa^{(a)}$ and $\kappa^{(b)}$ in terms of the polarization averaged absorption coefficient, $\kappa$ (Equation~\ref{avabsco}),
so that
\begin{eqnarray}
\kappa^{(a)} &=& \kappa\left(\frac{2\gamma+16/3}{\gamma+10/3}\sin^4\chi_H + \frac{4/3}{\gamma+10/3}\cos^4\chi_H 
                              +\frac{1}{2}\sin^{2}2\chi_H \right) \nonumber \\
            &=& \kappa \left(\frac{4/3 + 2(\gamma+2)\sin^2\chi_H}{\gamma+10/3}\right), \label{kapasimp} \\
\kappa^{(b)} &=& \kappa \left(\frac{2\gamma+16/3}{\gamma+10/3}\cos^4\chi_H + \frac{4/3}{\gamma+10/3}\sin^4\chi_H 
                              +\frac{1}{2}\sin^{2}2\chi_H \right) \nonumber \\
            &=& \kappa \left(\frac{4/3 + 2(\gamma+2)\cos^2\chi_H}{\gamma+10/3}\right), \label{kapbsimp}
\end{eqnarray} 
and        
\begin{eqnarray}
\kappa &=& D\mathcal{N}_0(K)B_0^{(2+\gamma)/2}\nu^{-(4+\gamma)/2}(M+N)^{(2+\gamma)/4} \label{kapsimp}
\end{eqnarray}
The values of $\kappa^{(a)}$ and $\kappa^{(b)}$ averaged over the initial magnetic field direction are thus
\begin{eqnarray}
<\kappa^{(a),(b)}> &=& D\mathcal{N}_0(K)B_0^{(2+\gamma)/2}\nu^{-(4+\gamma)/2} F_{\gamma+1}^{(a),(b)}(\delta,K) \label{kapabav} \\
<\kappa> &=& D\mathcal{N}_0(K)B_0^{(2+\gamma)/2}\nu^{-(4+\gamma)/2} H_\gamma(\delta,K) \label{kaptav}
\end{eqnarray}
where
\begin{eqnarray}
H_\gamma(\delta,K) &=& \frac{\int_{-\pi}^{\pi}\int_{0}^{\pi}(M+N)^{(2+\gamma)/4}
                                    \sin\theta d\theta d\phi}{4\pi}  \label{hdef} 
\end{eqnarray}

Unfortunately, the integrals appearing above are not, in general, analytically tractable. However, it is straightforward to evaluate
$F_\gamma^{(a),(b)}$ if $\gamma=3$ and $H_\gamma^{(a),(b)}$ if $\gamma=2$.  The results are 
\begin{eqnarray}
F_{\gamma=3}^{(a)}(\delta,K) &=& \frac{7+K^2 + \sin^2\delta(1-K^2)}{12K^2}  \label{kapa2} \\
F_{\gamma=3}^{(b)}(\delta,K) &=& \frac{7K^2+1 + 7\sin^2\delta(1-K^2)}{12K^2}  \label{kapb2} \\
H_{\gamma=2}(\delta,K) &=& \frac{1+K^2 +\sin^2\delta(1-K^2)}{3K^2}  \label{kapav2}
\end{eqnarray}
These results allow analytical calculation of the emission coefficients if $\gamma=3$ or the absorption coefficients if $\gamma=2$.

\begin{figure}
\begin{center}
\includegraphics[angle=-90,width=13.8cm]
{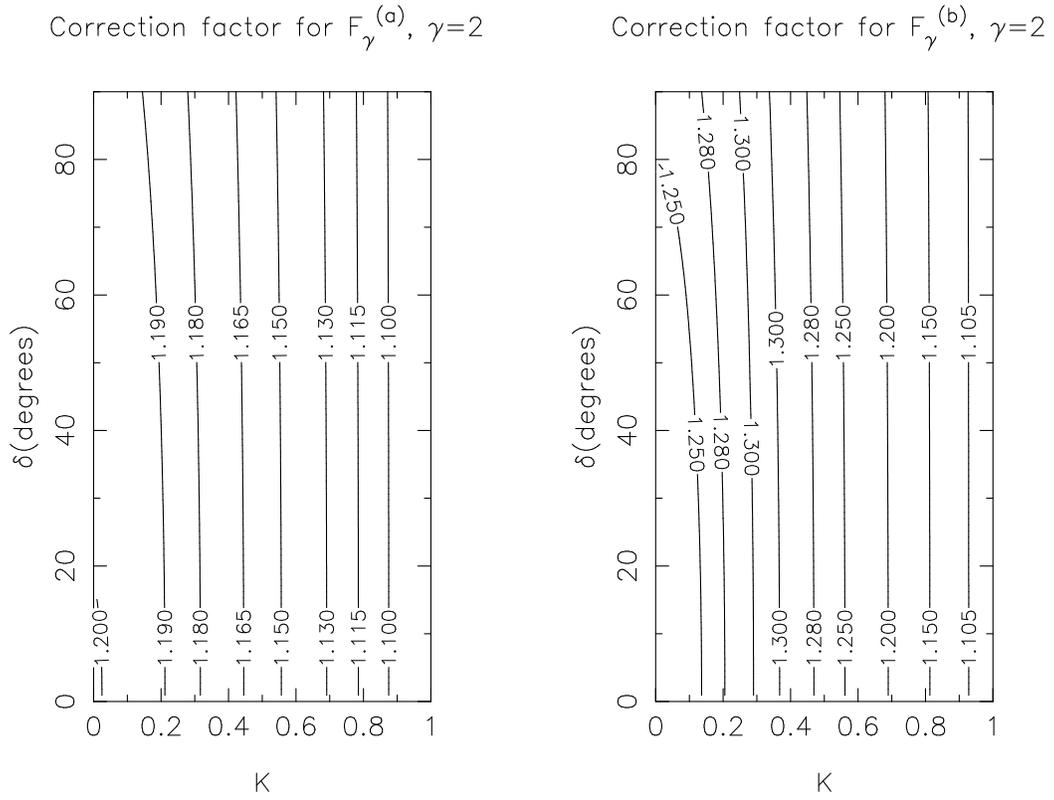}
\caption {This figure shows Factors for the correction of the approximate forms for $F^{(a)}$
and $F^{(b)}$ given in Equations~\ref{faapp} and \ref{fbapp}. \label{corf} }
\end{center}
\end{figure}

In an attempt to provide a means of calculating intensities without the aid of a computer 
various approximate solutions to the integrals for the $F$ and $H$ functions
were attempted. The more sophisticated approaches, such as rational
function approximations, were not successful. The best results overall
were obtained by setting $\gamma=2$ and making the rather crude approximation
that $(K^2(M+N))^{1/4} \simeq 1$ in Equations~\ref{edef} and \ref{fdef}. 
In that case,
\begin{eqnarray}
F_{\gamma=2}^{(a)}(\delta,K) \simeq f^{(a)}(\delta,K) &=& \frac{2}{39K^{(3/2)}}
                                                       \left(11 + 2K^2 + 2\sin^2\delta(1-K^2)\right) \label{faapp} \\
F_{\gamma=2}^{(b)}(\delta,K) \simeq f^{(b)}(\delta,K) &=& \frac{2}{39K^{(3/2)}}
                                                     \left(2 + 11K^2 + 11\sin^2\delta(1-K^2)\right) \label{fbapp}
\end{eqnarray}
The approximation for $F^{(a)}$ are accurate to within $20\%$, while that for $F^{(b)}$ is accurate to 
within $30\%$. While this is not very helpful by itself,  suitable correction factors, $T^{(a),(b)}$, where
\begin{eqnarray}
F_{\gamma=2}^{(a),(b)}(\delta,K) &=& f^{(a),(b)}(\delta,K) \times T^{(a),(b)}(\delta,K) 
\end{eqnarray}
are plotted in Fig.~\ref{corf}. In combination with Equations~\ref{kapa2} and \ref{kapb2}, these results
allow intensities and degrees of polarization to be determined for energy index $\gamma=2$, 
without resort to a computer. 

\section{Constants.}

Pacholczyk's treatment of synchrotron radiation in a uniform magnetic field involves a large number of physical constants
which are helpful in the derivations he performs. However, here, it is the constants
of proportionality, $C$ and $D$, (which are actually functions of $\gamma$) that are of 
greatest interest and expressions for them are given here. Additionally, the formulae
are converted from the obsolete CGS system to SI. 

Substituting expressions for the constants $c_1$, $c_3$ and $c_5$ into Equation 3.49
from Pacholczyk (1970) yields an expression for the emission coefficient in CGS units:
\begin{eqnarray}
\epsilon^{(1),(2)}_{CGS} = \frac{\beta}{16\sqrt{3}} \left(
1 \pm \frac{\gamma+1}{\gamma+7/3} \right) \frac{e^2}{c} 
       \left(\frac{3e}{2\pi mc}\right)^{(1+\gamma)/2} (mc^2)^{-(\gamma-1)}\mathcal{N}_0 
       H_{\perp}^{(1+\gamma)/2} \nu^{(1-\gamma)/2} \label{epscgs}
\end{eqnarray}
where, $-e$ is the electron charge, $m$ is the electron mass, $c$ is the speed of light in 
free space, $H_{\perp}$ is the component of magnetic field intensity perpendicular to the line of sight, and  
the numerical term $\beta$ is given by
\begin{eqnarray}
\beta &=& \Gamma\left(\frac{3\gamma-1}{12}\right) \Gamma\left(\frac{3\gamma+7}{12}\right)
          \frac{\gamma+7/3}{\gamma+1}
            \label{betadef}
\end{eqnarray}
The plus sign in Equation~\ref{epscgs} refers to polarization $1$ ($E$ perpendicular to the magnetic field)
and the minus sign to
polarization $2$ ($E$ parallel to the field). It is now straightforward to convert this expression to SI, resulting
in the formula
\begin{eqnarray}
\epsilon^{(1),(2)} = \frac{\beta}{16\sqrt{3}} \left(
1 \pm \frac{\gamma+1}{\gamma+7/3} \right) \frac{e^2}{4\pi\epsilon_0 c} 
       \left(\frac{3e}{2\pi m}\right)^{(1+\gamma)/2} (mc^2)^{-(\gamma-1)}\mathcal{N}_0 
       B_{\perp}^{(1+\gamma)/2} \nu^{(1-\gamma)/2} \label{epssi}
\end{eqnarray}
where $\epsilon_0$ is the permittivity of free space and $B_{\perp}=(B_a^2 + B_b^2)^{1/2}$. 
It follows that the constant $C$ in all expressions for the emission coefficients is given by
\begin{eqnarray}
C(\gamma) &=&  \frac{\beta}{16\sqrt{3}}
\frac{e^2}{4\pi\epsilon_0 c} 
       \left(\frac{3e}{2\pi m}\right)^{(1+\gamma)/2} (mc^2)^{-(\gamma-1)} \label{cdef}
\end{eqnarray}

Similarly, substituting for $c_6$ and $c_1$ in Equation 3.51 from Pacholczyk (1970) yields an expression
for the absorption coefficients
\begin{eqnarray}
\kappa_{CGS}^{(1),(2)} = \frac{\alpha}{16\sqrt{3}}\left(1 \pm \frac{\gamma+2}{\gamma+10/3}\right)\frac{e^2}{mc}
\left(\frac{3e}{2\pi mc}\right)^{(\gamma+2)/2}
        (mc^2)^{-(\gamma-1)} \mathcal{N}_0  H_{\perp}^{(2+\gamma)/2} \nu^{-(\gamma+4)/2} 
        \label{kappacgs}
\end{eqnarray}
where
\begin{eqnarray}
\alpha &=& (\gamma+10/3)\Gamma\left(\frac{3\gamma+2}{12}\right)\Gamma\left(\frac{3\gamma+10}{12} \right) \label{alphadef}
\end{eqnarray}
Again, this is easily converted to SI, giving
\begin{eqnarray}
\kappa^{(1),(2)} = \frac{\alpha}{16\sqrt{3}}\frac{e^2}{4\pi\epsilon_0 mc}\left(1 \pm \frac{\gamma+2}{\gamma+10/3}\right)
\left(\frac{3e}{2\pi m}\right)^{(\gamma+2)/2}
        (mc^2)^{-(\gamma-1)} \mathcal{N}_0  B_{\perp}^{(2+\gamma)/2} \nu^{-(\gamma+4)/2} 
        \label{kappasi}
\end{eqnarray}
It follows (by comparison with Equation~\ref{absco}) that the constant $D$ in the expressions for the absorption 
coefficients is given by
\begin{eqnarray}
D(\gamma) = \frac{\alpha}{16\sqrt{3}}\frac{e^2}{4\pi\epsilon_0 mc}\left(\frac{3e}{2\pi m}\right)^{(\gamma+2)/2}
        (mc^2)^{1-\gamma}
\end{eqnarray}

One further constant of importance appears in the term $\epsilon/ \kappa$, which appears when the expressions 
for the emission and absorption
coefficients are substituted into the uniform source solutions,~Equations~\ref{iasol} and \ref{ibsol}.
\begin{eqnarray}
\left(\frac{\epsilon}{\kappa}\right)^{(a),(b)} &=& \frac{C\mathcal{N}_0(K)B_0^{(1+\gamma)/2}\nu^{-(\gamma-1)/2}}{D\mathcal{N}_0(K)
B_0^{(2+\gamma)/2}\nu^{-(\gamma+4)/2}} \left(\frac{F_{\gamma}}{F_{\gamma+1}}\right)^{(a),(b)}   \nonumber \\
         &=& \mu \frac{m\nu^{5/2}}{\nu_L^{1/2}} \left(\frac{F_{\gamma}}{F_{\gamma+1}}\right)^{(a),(b)} \label{epskap}
\end{eqnarray}
where $\nu_L = eB_0/(2\pi m)$ is the cyclotron frequency in magnetic field $B_0$ and $\mu$ is the numerical value given by
\begin{eqnarray}
\mu &=&  \frac{ \Gamma\left(\frac{3\gamma-1}{12}\right)\Gamma\left(\frac{3\gamma+7}{12}\right)
            (\gamma+7/3)} 
                             {\sqrt{3} \Gamma\left(\frac{3\gamma+2}{12}\right)\Gamma\left(\frac{3\gamma+10}{12}\right)
            (\gamma+10/3)(\gamma+1)  } 
\end{eqnarray}

Numerical values of $\alpha$, $\beta$ and $\mu$ are given for common values of $\gamma$ in Table A1. 

\begin{table*}
\begin{center}
\caption{Numerical values for $\alpha$, $\beta$ and $\mu$.} 
\begin{tabular}{llll}
\tableline\tableline
    $\gamma$ & $\beta$ & $\alpha$ & $\mu$   \\
\tableline
       1.5   & 4.847   & 7.261 & 0.385 \\
       2.0   & 2.945   & 6.449 & 0.264 \\
       2.5   & 2.074   & 6.063 & 0.198 \\
       3.0   & 1.612   & 5.961 & 0.156 \\
       3.5   & 1.347   & 6.081 & 0.128 \\
\tableline
\end{tabular}
\end{center}
\end{table*}

\section{Variation of $\Pi_{thick}$ with $K$ in a simple model.}
\newpage
This section presents a magnetic field model that is similar to, but simpler than, that discussed in the 
main text of this paper. The aim is to illustrate more clearly the origin of the unusual behaviour
of $\Pi_{thick}$  shown in Fig.~\ref{pivk} (lower panel) in which, as $K$ increases from zero,
(reducing the anisotropy of the magnetic field) then for $\delta=0$, $\Pi_{thick}$ actually increases.
This behaviour is more easily understood in the case of a  
source in which the magnetic field is in the plane of the sky. It consists of a large number of cells,
a fraction $(1-x)$ of which
have magnetic field $B$ parallel to the $b$ axis with value $B_0/K$, and a fraction $x$
have $B$ parallel to the $a$ axis with value $B_0$ ($K<1$).
The fact that this field configuration doesn't satisfy ${\bf{\nabla.B}}=0$ does not detract from its usefulness
for the present purpose. 

Since the magnetic field is in the sky plane in one of two orthogonal directions, Equations~\ref{emco} and \ref{absco}
give the emission and absorption coefficients for the fraction $(1-x)$ of cells with  
$B$ parallel to the $b$ axis as 
\begin{eqnarray}
\epsilon^{(a),(b)} &=& CN(B_0/K)^{3/2}(1 \pm s) \label{expepsa}  \\
\kappa^{(a)(,b)} &=& DN(B_0/K)^2(1 \pm r)       \label{expkapa}           
\end{eqnarray}
where, for $\gamma=2$,
$s = (\gamma+1)/(\gamma+7/3)=9/13$,  
$r = (\gamma+2)/(\gamma+10/3)=3/4$  
and the upper and lower symbols in the plus or minus sign refer to polarizations $(a)$ and $(b)$ respectively.
For the fraction $x$ of cells with magnetic field parallel to the $a$ axis, the $(1+r)$ factors become $(1-r)$
and vice versa, and similarly for $(1\pm s)$. The magnetic field becomes $B_0$.
For these cells, the emission and absorption coefficients are therefore
\begin{eqnarray}
\epsilon^{(a),(b)} &=& CN(B_0)^{3/2}(1 \mp s) \label{expepsa2} \\
\kappa^{(a),(b)} &=& DN(B_0)^2(1 \mp r)       \label{expkapa2}  
\end{eqnarray}
The total contribution to $\epsilon^{(a)}$ is therefore
\begin{eqnarray}
\epsilon^{(a)} &=& CN_0(B_0/K)^{3/2}((1-x)(1+s) + K^{3/2}x(1-s)) \nonumber \\
             &=&  CN_0(B_0/K)^{3/2}(1-x)(1+s)\left(1+
                  \frac{x}{1-x}\frac{1-s}{1+s}K^{3/2} \right)  \label{epsatot}
\end{eqnarray}
Similarly, the remaining coefficients are
\begin{eqnarray}
\epsilon^{(b)} &=& CN_0(B_0/K)^{3/2}(1-x)(1-s)\left(1+
             \frac{x}{1-x}\frac{1+s}{1-s}K^{3/2}\right) \label{epsbtot}  \\
\kappa^{(a)} &=& DN_0(B_0/K)^{2}(1-x)(1+r) \left(1+ 
           \frac{x}{1-x}\frac{1-r}{1+r}K^2\right) \label{kapatot} \\
\kappa^{(b)} &=& DN_0(B_0/K)^2(1-x)(1-r) \left(1+ 
           \frac{x}{1-x}\frac{1+r}{1-r}K^2\right) \label{kapbtot}
\end{eqnarray}
In the optically thick limit, the degree of polarization is
\begin{eqnarray}
\Pi_{thick} &=& \frac{(\epsilon^{(a)}/ \kappa^{(a)}) - (\epsilon^{(b)}/ \kappa^{(b)})}
        {(\epsilon^{(a)}/ \kappa^{(a)}) + (\epsilon^{(b)}/ \kappa^{(b)})}
          = \frac{Q-1}{Q+1}   \label{pithq}
\end{eqnarray}
where
$Q = (\epsilon^{(a)}\kappa^{(b)})/(\epsilon^{(b)}\kappa^{(a)})$. 
If $K\rightarrow 0$, $\Pi_{thick}$ is as for a uniform field, i.e. negative, and $0<Q<1$. 
Then, if $Q$ increases with increasing $K$, $|\Pi_{thick}|$ decreases. If
$Q$ decreases with increasing $K$, then $|\Pi_{thick}|$ increases. 
Substituting from Equations,~\ref{epsatot} to \ref{pithq}, 
\begin{eqnarray}
Q &=& \frac{1+s}{1-s}\frac{1-r}{1+r}\frac{1+XSK^{3/2}}{1+XS^{-1}K^{3/2}}\frac{1+XR^{-1
}K^2}{1+XRK^2} \label{qexp}
\end{eqnarray}
where
$S = (1-s)/(1+s)$, 
$R = (1-r)/(1+r)$, and
$X = x/(1-x)$.
If $K\ll1$, then, neglecting terms of order $K^3$ and higher
\begin{eqnarray}
Q & \simeq & \frac{R}{S}\left(1+X(S-S^{-1})K^{3/2} + X(R^{-1}-R)K^2\right) \nonumber  \\ 
  & \simeq & \frac{R}{S}\left(1+X\left(\frac{-4sK^{3/2}}{1-s^2} + \frac{4rK^2}{1-r^2}\right)\right) \label{qfin}
\end{eqnarray}
which will decrease with increasing $K$ if $dQ/dK < 0$, i.e. if
\begin{eqnarray} 
\frac{3sK^{1/2}}{2(1-s^2)} &>& \frac{2rK}{1-r^2}  \label{cond1}
\end{eqnarray}
or
\begin{eqnarray}
K &<& \frac{9}{16}\left(\frac{s}{1-s^2}\frac{1-r^2}{r}\right)^2 = 0.34 \label{cond2}
\end{eqnarray}
to two significant figures, if $\gamma=2$.
So provided $K$ is very small, as $K$ increases, $Q$ decreases and $|\Pi_{thick}|$ increases,
while $\Pi_{thick}$ is negative. This happens because as $K$ increases, the emission process tends
to favour $I^{(b)}$ over $I^{(a)}$ (i.e. $\epsilon^{(b)}$ increases more than $\epsilon^{(a)}$). However, 
the absorption process (or the mean free path) favours $I^{(a)}$
over $I^{(b)}$ (because $\kappa^{(b)}$ increases more than $\kappa^{(a)}$). If $K$ is small, the effect on $Q$
due to the change in emission coefficients ($\propto K^{3/2}$) dominates that due to the change in absorption 
coefficients because ($\propto K^{2}$) when $K \ll 1$.

Comparison with of Inequality~\ref{cond2} with the position of the turning point on the $\delta=0$ curve 
from the lower panel of Fig.~\ref{pivk},
shows that, in the compressed random field model, $|\Pi_{thick}|$ increases with $K$ over a more limited
range of $K$, from $K=0$ to about $0.2$, rather than $0.34$.
This discrepancy arises because, in the model compressed random field model, when $K$ is small, 
the field parallel to the $b$ axis is like a plate of spaghetti, much of it points toward
us, reducing the emission coefficients of this component by $\sin^{3/2}\theta$ and the absorption coefficients
by $\sin^{2}\theta$, where $\theta$ is the inclination of the field to the line of sight. 
The result is to replace $K$ in the above expressions by $K/ \sin\theta$. This will tend to make the condition 
on $K$ more stringent than given by Inequality~\ref{cond2}.

\end{document}